\documentclass[apj]{emulateapj}
\newcommand\submitms{n}		
\if\submitms y
\else
\usepackage{apjfonts}
\fi

\usepackage{ifthen}
\usepackage{natbib}

\newcommand\degree{\degr}
\newcommand\degrees\degree
\DeclareSymbolFont{UPM}{U}{eur}{m}{n}
\DeclareMathSymbol{\umu}{0}{UPM}{"16}
\let\oldumu=\umu
\renewcommand\umu{\ifmmode\oldumu\else$\oldumu$\fi}
\newcommand\micro{\umu}
\renewcommand\micron{\micro m}
\newcommand\microns \micron

\newcount\timect
\newcount\hourct
\newcount\minct
\newcommand\now{\timect=\time \divide\timect by 60
         \hourct=\timect \multiply\hourct by 60
         \minct=\time \advance\minct by -\hourct
         \number\timect:\ifnum \minct < 10 0\fi\number\minct}
\newcommand\vs{\em vs.}

\shorttitle{SL9 Impact Models: 3D Bolide}
\shortauthors{Korycansky Harrington Deming Kulick}

\slugcomment{{\em ApJ}, in press.}
\received{2005 November 7}
\revised{2006 March 21}
\accepted{2006 March 30}

\begin{document}

\title{Shoemaker-Levy 9 Impact Modeling: I. High-Resolution 3D Bolides}
\if\submitms y
  \author{D.\ G.\ Korycansky}
  \affil{CODEP, Department of Earth Sciences, \\
         University of California, Santa Cruz CA 95064}
  \email{kory@pmc.ucsc.edu}
  \and
  \author{Joseph Harrington}
  \affil{326 Space Sciences Building, Center for Radiophysics and Space
               Research \\
         Cornell University, Ithaca, NY 14853-6801}
  \email{jh@oobleck.astro.cornell.edu}
  \and
  \author{Drake Deming}
  \affil{NASA/Goddard Space Flight Center,
               Planetary Systems Branch, Code 693 Greenbelt, MD 20771}
  \email{ddeming@pop600.gsfc.nasa.gov}
  \and
  \author{Matthew E.\ Kulick}
  \email{mek28@cornell.edu}
  \affil{Center for Radiophysics and Space Research \\
               Cornell University, Ithaca, NY 14853-6801}
\else
  \author{
    D.\ G.\ Korycansky\altaffilmark{1},
    Joseph Harrington\altaffilmark{2},
    Drake Deming\altaffilmark{3},
    and
    Matthew E.\ Kulick\altaffilmark{2}
  }
  \altaffiltext{1}{CODEP, Department of Earth Sciences,
        University of California, Santa Cruz CA 95064; kory@pmc.ucsc.edu.}
  \altaffiltext{2}{Center for Radiophysics and Space Research,
        Cornell University, Ithaca, NY 14853-6801 USA;
         jh@oobleck.astro.cornell.edu, mek28@cornell.edu.}
  \altaffiltext{3}{Planetary Systems Branch, NASA's Goddard Space
	Flight Center, Greenbelt, MD 20771-0001; ddeming@pop600.gsfc.nasa.gov.}
\fi

\begin{abstract}


We have run high-resolution, three-dimensional, hydrodynamic simulations
of the impact of comet Shoemaker-Levy 9 into the atmosphere of Jupiter.  We 
find that the energy deposition profile is largely similar to the previous 
two-dimensional calculations of \citeauthor{MacLowZahnle1994apj}, though 
perhaps somewhat broader in the range of height over which the energy is 
deposited.  As with similar calculations for impacts into the Venusian 
atmosphere, there is considerable sensitivity in the results to small changes 
in the initial conditions, indicating dynamical chaos.  We calculated the 
median depth of energy deposition (the height $z$ at which 50\% of the 
bolide's energy has been released) per run.  The mean value among runs is 
$\approx$ 70 km below the 1-bar level, for a 1-km diameter impactor of porous 
ice of density  $\rho=0.6$ g cm$^{-3}$.  The standard deviation among these 
runs is 14 km.  We find little evidence of a trend in these results with the 
resolution of the calculations (up to 57 cells across the impactor radius, or 
8.8-m resolution), suggesting that resolutions as low as 16 grid cells across 
the radius of the bolide may yield good results for this particular quantity.

Visualization of the bolide breakup shows that the ice impactors were shredded 
and/or compressed in a complicated manner but evidently did not fragment into 
separate, coherent masses, unlike calculations for basalt impactors.  The 
processes that destroy the impactor take place at significantly shallower 
levels in the atmosphere ($\sim -40$ km for a 1-km diameter bolide) but the 
shredded remains have enough inertia to carry them down another scale height or
more before they lose their kinetic energy.

Comparion basalt-impactor models shows that energy deposition
curves for these objects have much less sensitivity to initial conditions
than do ice impactors, which may reflect differences in the equation
of state for the different kinds of objects, or a scale-dependent
breakup phenomenology, with the preferred scale depending on impactor
density.

Models of impactors covering a $\sim 600$-fold range of mass ($m$)
show that larger impactors descend slightly deeper than expected from
scaling the intercepted atmospheric column mass by the impactor mass.
Instead, the intercepted column mass scales as $m^{1.2}$.

\end{abstract}


\section{INTRODUCTION}
\label{intro}

For the week starting 16 July 1994, fragments of comet Shoemaker-Levy
9 (SL9) hit Jupiter.  Most of the world's telescopes observed the
event, collecting an unprecedented volume of imaging, photometry, and
spectroscopy that spanned all sensible wavelengths.  Papers in
collections edited by
\citet{WestBoenhardt1995bkesosl9} and \citet{NollEtal1996bksl9iau156} review 
the data and early interpretation.  \citet{HarringtonEtal2004jupbksl9}
review the phenomenology, efforts to understand the impact phenomena,
and open questions about the impacts.  A brief summary of points
relevant to this paper follows.

The impacts followed the same basic phenomenology.  The orbital path of the 
comet fragments intersected the planetary surface at $\sim$45{\degrees} S 
latitude, and planetary rotation arranged for a girdle of well-separated 
impacts there.  Each impactor fell into the atmosphere at over 60 km s$^{-1}$ 
and an impact angle of about 45{\degrees} \citep{ChodasYeomans1996}.  The 
ground track of the impactors moved toward the northwest.  The bolides 
crushed, ablated, and decelerated as they fell through the atmosphere, leaving 
an entry channel filled with superheated gas \citep[e.g.,][]
{AhrensEtal1994grlsl9imp, AhrensEtal1994grlsl9rad,
BosloughEtal1995grlsl9, BosloughEtal1997grldayglow, Crawford1996,
CrawfordEtal1994swsl9, MacLow1996, MacLowZahnle1994apj,
TakataEtak1994icsl9, TakataAhrens1997icsl9,ShuvalovEtal1999ijie}.  
This gas rushed back out the 
entry channel, exiting the atmosphere at an angle and flying ballistically 
into space.  The Hubble Space Telescope (HST) resolved the plumes, and saw 
several rise to $\sim$3000 km above the cloud tops
\citep{HammelEtal1995scisl9}. 
Material invisible from Earth rose higher still, as shown by its
effects on the magnetosphere \citep{BrechtEtal2001icsl9,
BrechtEtal1995grlsl9}.  The plumes collapsed and
re-entered the atmosphere in 20 minutes, heating it and leading to infrared 
emission so strong that it was dubbed the ``main event''.  In the initial hours
after impact, two different systems of expanding rings were seen, one in the 
visible by HST \citep{HammelEtal1995scisl9} and the other in the infrared by
\citet{McgregorEtal1996icsl9ring}.  Peculiar patterns remained in the 
atmosphere, which were spread in latitude over the ensuing weeks 
\citep{BanfieldEtal1996icsl9}.

Chemistry was just as exciting.  Main-event spectra were complex, but
only five  
species were identified.  Dozens appeared in the aftermath, and CO and CO$_2$ 
are observed to this day, now having crossed the equator into the northern 
hemisphere \citep{BezardEtal2002icsl9co, LellouchEtal2002icsl9co2h20}.  Not to 
be outdone, the magnetosphere responded strongly and in a manner that changed 
throughout the week of impacts as ionized plume material loaded Jupiter's
magnetic field lines.

Initially viewed as perfect perturbation experiments, the impacts instead 
involved complex phenomena at fine spatial and temporal timescales.  Models 
capable of reproducing many of the observed phenomena were too complex by 
several orders of magnitude for the computational power available at the time. 
Many observers, including most spectroscopists, still have unpublished data 
that they cannot interpret.

We made some headway in explaining phenomena of the images and
lightcurves with simplified calculations:
\citet{HarringtonDeming2001apjsl9i} extrapolated a published blowout
velocity distribution to calculate the fluxes of mass, energy, and momentum on 
the atmosphere.  \citet{DemingHarrington2001apjsl9ii} used those fluxes to 
drive a two-dimensional (2D) vertical-slice model of the atmosphere.  The 
resulting lightcurves are a good match to data from 0.9 -- 12 {\microns}, and 
the time-dependent pressure ($p$) and temperature ($T$) grids show phenomena 
that mimic McGregor's Ring and other effects.

Encouraged by this initial success, we have undertaken a study to model the 
impacts and their aftermath sufficiently to reproduce, in a realistic manner, 
all of the observed phenomena in the impact, blowout, plume flight, and plume 
splash phases.  By ``realistic'', we mean that wherever we can eliminate an 
ad-hoc assumption or an analytic approximation, we do.  We no longer use 
arbitrary initial conditions.  Rather, we will use observations and the 
results of prior models for initialization, and will ultimately
produce synthetic images, 
lightcurves, and spectra derived with radiative transfer from model results.  
For physical atmospheric effects, our approach is the chaining of hydrodynamic 
models outlined by \citet{HarringtonEtal2004jupbksl9}.  Our chemical models 
will be driven by tracer particle histories from the physical models.

This paper presents the first results from our impact model.  Since the 
observations of this phase were not as constraining as those of later phases, 
our primary goal was to produce a data grid with which to initialize subsequent 
models.  Doing this believably required a look at, e.g., resolution convergence
and sensitivity to initial conditions.  Also, and as with any interesting 
investigation, there are serendipitous findings that reach beyond this 
particular set of impacts.

\citet{KorycanskyEtal2002icar} and \citet{KorycanskyZahnle2003icar} have made
3D calculations of the impact of asteroids into the atmosphere of Venus using 
ZEUS-MP and its predecessor ZEUS3D. There are a number of similarities between
the results of that work and the present study that we note below.  
Comparable hydrodynamical simulations of the initial phase of the impacts
have been previously described by a number of groups \citep{BosloughEtal1994grl,
CrawfordEtal1994swsl9, GryaznovEtal1994emp,TakataEtak1994icsl9,
YabeEtal1994,ShoemakerEtal1995grl, CrawfordEtal1995ijie,
Svetsov1995sola,YabeEtal1995grl,Crawford1997nyacad,ShuvalovEtal1997sola,
ShuvalovEtal1999ijie}.

Section 2 presents our model.  We describe the results of over a dozen runs 
in section 3, and discuss their implications and future work in section 4.

\section{IMPACT MODEL}
\label{method}
For the calculations described in this paper we employed the ZEUS-MP
hydrodynamics code \citep{Norman2000}, which solves the equations for
three-dimensional compressible gas flow. 
We have modified ZEUS-MP (base version 1.0) to include 
multiple materials. Modifications to the code are described in more detail by 
\citet{KorycanskyEtal2002icar} and \citet{KorycanskyZahnle2003icar}. 
We have not included radiative transfer in ZEUS-MP; the short timescale
and high optical depth of the atmosphere below 1 bar, where the main
disruption of the impactors and energy deposition takes place, makes it unlikely
that radiative transfer would significantly affect the dynamics. That
assumption has been tested by \citet{ShuvalovEtal1999ijie}, who indeed found
that dynamical effects of radiation transfer were insignificant, but that
the impacting comets would be strongly heated at large depths.

The Jovian atmospheric profile comes from  
\citet{DemingHarrington2001apjsl9ii}.
We also included minor modifications such as a moving
grid that tracks the impactor at a variable velocity as it
decelerates.  Multiple materials were handled by the integration of tracer
variables advected in the flow.
For non-porous material, the tracer $C$ gives the fraction of mass in
the cell that is impactor material.
Porosity is tracked by an additional tracer, and is treated with the
so-called $p-\alpha$ model for a strengthless, porous solid 
\citep{MenikoffKober1999}.  The coordinate system ($x_1$, $x_2$, $x_3$)
is Cartesian and aligned with the bolide's initial velocity such that
$x_1$ is the along-track coordinate, $x_2$ is horizontal, and $x_3$ is
perpendicular to
the others.  We relate ($x_1,x_2,x_3$) to local
Cartesian coordinates ($x,y,z$) as follows:
\begin{eqnarray}
x &=& x_2  \\
y &=& -x_1\sin\theta + x_3\cos\theta  \\
z &=& x_1\cos\theta + x_3\sin\theta, \label{coord}
\end{eqnarray}
where $\theta$ is the angle between $x_1$ and the vertical.  Note that
local coordinates are not cardinal directions.
Fluid velocities in the $x_1$, $x_2$, and $x_3$ directions are 
$v_1$, $v_2$, and $v_3$, respectively.  Other quantities that appear
in the equations
are the density, $\rho$ and the internal energy per unit volume, $e$. 
The spatial resolution of the calculations
is described by the notation R$n$, where $n$ is the number of grid cells
across the radius of the body in the high-resolution part of the grid.
Away from an inner block of dimensions $4\times 2 \times 2$ km, 
the grid spacing increases geometrically (by a factor $\sim 1.04$
per grid cell, depending on the overall resolution).
The computational
grid moves with the impactor and decelerates, keeping the object's
front end about 1 km from the front end of the grid so that the object
remains in the high-resolution portion as it disrupts.
Calculations in the paper were made with resolutions of R16, R32, and R57.

We used the Tillotson equation of state, which was formulated for high-velocity
impacts \citep{Tillotson1962,Melosh1989}, although it cannot represent
melting or mixed two-phase (gas-liquid) regimes.
The Tillotson EOS has two regimes, one for cold and/or
compressed material, the other for rarefied, hot conditions.  
For a given mass density of impactor material $\rho$
and internal energy density (energy per unit volume) $e$,
we have ${\cal E}=e/\rho$, $\eta=\rho /\rho_T$, $q=\eta-1$, where $\rho_T$
is the density at zero pressure and ${\cal E}$ is the specific energy.  Then, 
for compressed states ($\eta>1$) and expanded ones ($\eta<1$) where
${\cal E}<{\cal E}_{iv}$, the incipient vaporization energy, the pressure $P$
is the sum of a thermal term and a cold-pressure term:
\begin{equation}
P = P_l = \left(a+{b\over 1+{\cal E}/({\cal E}_T\eta^2)}\right)e+
{Aq+Bq^2\over 1+e^{-K_c q}},\label{eq:tilcold}
\end{equation}
The parameters $a$, $b$, $A$, $B$, and ${\cal E}_T$ are described in more
detail and listed for common substances by \citet{Melosh1989}; $K_c$ is
described below.  For expanded states for which ${\cal E}>{\cal E}_{cv}$, the 
energy of complete vaporization, 
\begin{equation}
P= P_h = ae+\left[{be\over 1+{\cal E}/({\cal E}_T\eta^2)}+{Aq\over 1+e^{-K_c q}}
\right]e^{-5(1/\eta -1)^2}. \label{eq:tilhot}
\end{equation}
We modify the
cold-pressure terms $Aq+Bq^2$, and $Aq$ in Eqs.\ (\ref{eq:tilcold}) and
(\ref{eq:tilhot}) respectively, by a term of the form $1+e^{-K_c q}$ in the
denominator, in order to provide a low-density cutoff as recommended by
\citet{Melosh1989}.  We set $K_c=1000$.   A strengthless
material is assumed, so negative pressures (tensions) are not sustainable;
the pressure cutoff for low densities mimics that effect.
For expanded states in which ${\cal E}_{iv} < {\cal E} < {\cal E}_{cv}$, $P$
is linearly interpolated between $P_l$ and $P_h$.  Then, on the grid,
the resulting Tillotson pressure is weighted by the tracer $C$.
The equation of state parameters for our materials (ice and basalt)
are the same as used by \citet{BenzAsphaug1999icar} and are listed in table I.
(As discussed below, we modeled basalt impactors
in order to study the effect of the equation of state on impactor breakup.)

\begin{table*}
\begin{center}
\caption{Tillotson EOS parameters}
\begin{tabular}{lcccccccc}
\hline
material & $\rho_T$ (g cm$^{-3})$ & $a$ & $b$ & $A$ (erg cm$^{-3}$) & $B$ 
(erg cm$^{-3}$) & 
${\cal E}_T$ (erg g$^{-1}$) & ${\cal E}_{iv}$(erg g$^{-1}$) & ${\cal E}_{cv}$ 
(erg g$^{-1}$) \\
\hline
ice    & 0.917 & 0.3 & 0.1 &  $9.47\times 10^{10}$ & $9.47\times 10^{10}$ & 
$1.0\times 10^{11}$ & $ 7.73\times 10^{9}$ & $3.04\times 10^{10}$ \\
basalt & 2.70 & 0.5 & 1.5 &  $2.67\times 10^{11}$ & $2.67\times 10^{11}$ & 
$4.87\times 10^{12}$ & $4.72\times 10^{10}$ & $1.82\times 10^{11}$ \\
\hline
\end{tabular}
\end{center}
\end{table*}

For porosity, we use a simple model
for a strengthless, porous solid based on the ``$p-\alpha$''
formulation
\citep{Herrmann1969, MenikoffKober1999}. 
We distinguish between the solid material
density $\rho_m$ (and energy $e_m$)
and the same quantities for fluid in the porous
void space ($\rho_m$ and $e_m$).  Space and material quantities
are related by the porosity $\epsilon$, where $0 < \epsilon < 1$:
$\rho=(1-\epsilon)\rho_m$ and
$e=(1-\epsilon)e_m$.  The material pressure $p_m$ is provided by
the equation of state $p_m(\rho_m,e_m)$ (the Tillotson EOS described above)
and the space pressure is thus $p=(1-\epsilon)p_m$.  Initially the porosity
is $1-$(initial density/nominal density) of the material.
During the calculation, porosity is
advected with the flow like the material tracer $C$.
At each timestep the advected value of $\epsilon$ is then adjusted by
comparison with a function $\epsilon_f(p,p_c)=\epsilon_0(1-p/p_c)^2$ that
depends on a ``crushing pressure'', $p_c$, which for ice we set to 
$10^7$ dyne cm$^{-2}$.  For $p>p_c$, $\epsilon_f=0$.
The porosity is given by $\epsilon={\rm min}(\epsilon,\epsilon_f)$.
During the calculation, porosity can only decrease from its initial value; 
if porosity is crushed out of a mass element, it does not return even if the 
pressure drops back below $p_c$.

The Tillotson EOS has been used for SL9 calculations by 
\citet{GryaznovEtal1994emp} and \citet{TakataEtak1994icsl9}. 
Other equations of state for cometary material used in previous calculations 
include perfect gas (with adiabatic indices $\gamma = 1.2$ and 1.4),
\citep{MacLow1996},  and a ``stiffened gas'' or Murnaghan EOS with an 
additional perfect-gas thermal contribution \citep{MacLowZahnle1994apj}.
A similar stiff equation of state was used by 
\citet{YabeEtal1994,YabeEtal1995grl}.
\citet{BosloughEtal1994grl}, 
\citet{CrawfordEtal1994swsl9,CrawfordEtal1995ijie}, and 
\citet{Crawford1997nyacad} used a tabular version of the ANEOS equation of
state that took into account melting and vaporization.
\citet{ShuvalovEtal1999ijie} used a combination of a non-linear Gruneisen
EOS in which the pressure increased roughly as the cube of the density
ratio $\rho/\rho_T$ combined with a tabular gas EOS for the vaporized
component.

The most important factors in the various equations of state are probably
the relative stiffness $dP/d\rho$ and
the presence or absence of a dependence of pressure on the internal energy
$e$.  A relatively stiff EOS, and one in which $P$ is also a function of $e$,
might be expected to result in calculations that show impactors blowing up
at greater altitudes and exhibiting more radial spreading (``pancaking'')
than otherwise.  This may explain some of the differences among the results
that have been found from different studies.  We will not systematically 
explore that question in this paper, but in view of the differences
we found (discussed below) in results between ice and basalt impactors, we
plan to address the issue in future work. 

Our ``standard case'' is a 1-km-diameter, porous ice sphere impacting the 
atmosphere at $v=61.46$ km $s^{-1}$ and $\theta = 43.09^{\circ}$.  The
velocity and impact angle are the means of those of the 21 comet fragments
\citep{ChodasYeomans1996} in a frame rotating with Jupiter at the average 
latitude of the impacts ($-44.02^{\circ}$).  The gravitational acceleration
at that latitude (including the J2 and centrifugal terms) is $g=2512$
cm s$^{-2}$.  The bulk density of the bolide is $\rho=0.6$ g cm$^{-3}$
\citep{AsphaugBenz1994natsl9, AsphaugBenz1996icsl9, Solem1994natsl9,
Solem1995aasl9}.

We have also carried out calculations of impacts with objects of 
bulk density $\rho=0.1$ and 0.92 g cm$^{-3}$, the latter corresponding to
non-porous solid ice.  In addition, we have done calculations with impactors
with volumes of 0.2, 3, 40, and 125 times that of the standard case,
or diameters of 0.584, 1.44, 3.42, and 5 km.  Not all combinations of 
density, diameters, and resolutions were run.

One important result that emerged from the Venus-atmosphere calculations
was the significant extent to which the results were sensitively 
dependent on initial conditions in a manner reminiscent of dynamical chaos 
\citep{KorycanskyZahnle2003icar}.   Two 
calculations whose initial conditions (for instance, impact velocity)
differed by very small amounts ($\sim 0.1$\%) would develop large divergences
in the course of their respective simulations.  \citet{KorycanskyEtal2000icar}
took the view that the basic process of impactor disruption was due to
the growth of Rayleigh-Taylor and Kelvin-Helmholtz instabilities
\citep{FieldFerrara1995apj}, and it is
plausible that the seeds of the perturbations that grow to saturation are
irregularities due to the finite resolution on the grid.  In the physical
case, one would expect analogous irregularities from the inevitable bumpiness
of the bolide's surface.  While the basic process
(fragmentation and ablation) was always the same, there
could be significant differences in the results of different runs for
impactors of the same gross properties. For example, the diameter of the 
resulting crater on the surface of Venus might vary by several kilometers
(on a scale of $\sim 10$ km) depending on the exact details of how the event 
had unfolded.  In the limit where the bolide just reached the surface,
different 
cases might result in anything from a single crater several km in diameter,
a group of small craters, or no crater at all.  We might expect
similar behavior to obtain in this case, and we have attempted to sample
the distribution of results by performing several runs of the standard
case with small differences in initial impact velocity $\Delta v$ of 0.1\% of 
the initial velocity $v$ ($\approx 0.06$ km s$^{-1}$) or displacements 
in the cross-track coordinates $\Delta x_2$, $\Delta x_3$ by one half of a 
grid cell. For R16 calculations of the 1-km bolide the displacements are 
15 m, for R32 they are 7.5 m, and for R57 they are 4.4 m.

We use several diagnostics to characterize the calculations.  The simplest
and most significant is the profile of deposition of impactor
kinetic energy $dE/dz$ as a function of height $z$ in the atmosphere,
the same quantity as discussed by \citet{MacLowZahnle1994apj} and
\citet{MacLow1996}.
The impactor kinetic energy $E=E(z)$ was calculated by integrating
the flux of kinetic energy passing through a given height $z$:
\begin{equation}\label{energy_flux}
E(z) = \int \! dt \int\int_z \! \! \rho C {(v_1^2+v_2^2+v_3^2)\over 2} 
v_1dx_2dx_3.\label{eflux}
\end{equation}
The area integral is taken over all cells whose height
$z=x_1\cos\theta+x_3\sin\theta$ is the desired value. [Note that the projection
factors $\cos\theta$ in the element of area in the $z$ plane
($dA=dx_2dx_3/\cos\theta$) and the downward velocity ($v_z=v_1\cos\theta$)
cancel.]   The time integral extends to the end of the
calculation (typically  8--10 seconds).
The density on the grid is weighted by the advected tracer field $C$
that tags material that belongs to the impactor.
We computed $E(z)$ for $-200 \leq z\leq 100$ km
at 1-km intervals. (For more massive bolides, $E(z)$ was calculated as far
down as needed.) Having integrated $E(z)$ for all heights, 
the deposition profile $dE/dz$ follows by numerical differentiation.

For a subset of the runs, we
calculated spatially resolved plots of mass flux, 
similar to those employed by \citet{KorycanskyZahnle2004icar} to
study the fragmentation of asteroids in the Venusian atmosphere.  Here we
hope to correlate events such as fragmentation, as revealed in the mass-flux
plot, to specific features (like peaks) in the deposition profile. 

The time-integrated mass flux $\mu(z;x,y)$ 
at a height $z$ is given by
\begin{equation}
\mu(z;x,y)=-\int \rho(z;x,y) C(z;x,y) v_1(z;x,y) dt,\label{mudef}
\end{equation}
where $z$
translates to the tilted plane $x_1\cos\theta+x_3\sin\theta$ in the
grid coordinate system.  In practice, $\mu$ is calculated as the accumulation
of a set of integrals at time slices $t_i$,
each integrated over the interval $t_i-\Delta t_i/2$ to $t_i+\Delta t_i/2$,  
by assuming that material moves at a constant velocity during that interval 
and counting all the mass that has crossed or will cross the plane 
during the interval.  Due to the tilt of the grid, the $y$-location of the 
impactor is a function of $z$; in this case we refer the position of 
material on the plane to the centerline position defined by $x_2=x_3=0$.

The calculations were performed on a Beowulf cluster using 32 2.4 GHz Opteron
processors.  The largest (R57) runs took $\sim 3.1\times 10^7$ cpu seconds on a 
grid of $712 \times 356 \times 356 = 9.0\times 10^7$ points.  Impactors of
1 km diameter took $\sim 8-10$ model seconds for the velocity to
fall to $0.001 \times$ the initial velocity, which was the condition for
stopping the calculation.
Timesteps were $\sim 5\times 10^{-5}$ s during the initial phases of the
impact, increasing to $\sim 3\times 10^{-4}$ s by the end, as the impactor
slowed down.  High-resolution output was written to disk every 0.1 model 
seconds, as noted above.  The R57 calculations occupied 17 GB of memory and 
produced $\sim 210$ GB of output data.  Not all data were saved from all runs.
Wall-clock time for an R57 run was about three weeks.

\section{RESULTS}
\label{results}

\begin{figure*}
\centerline{\includegraphics[width=14cm]{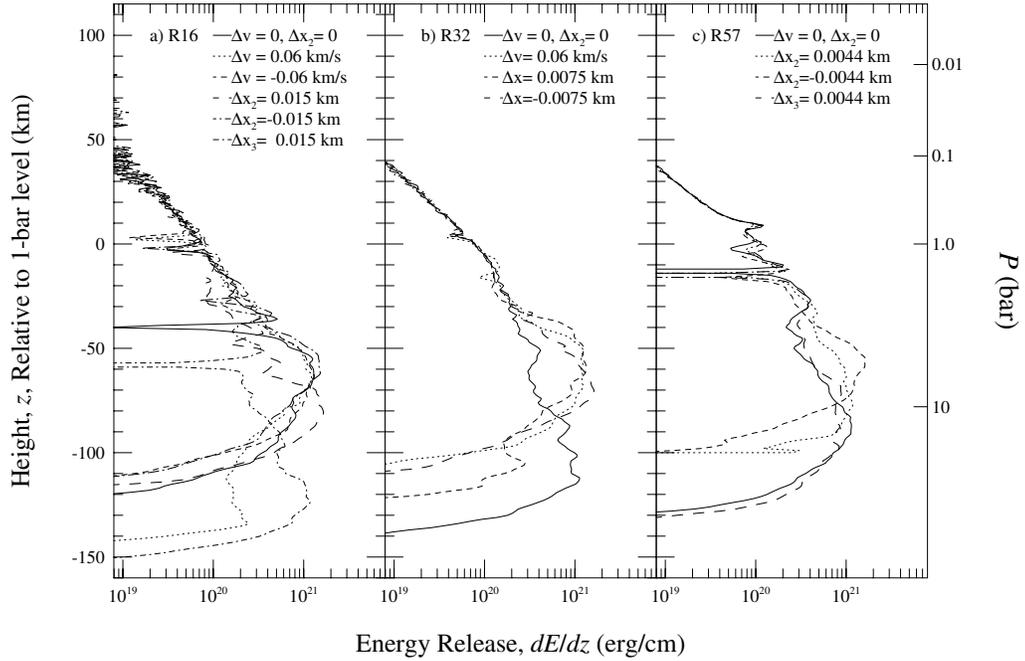}}
\caption{\label{figedeplog}
Energy-deposition profiles for realizations of the impact of a
1-km-diameter ice sphere ($\rho =$ 0.6 g cm$^{-3}$) into Jupiter's
atmosphere at 61.4 km s$^{-1}$.  Calculations at three different resolutions
(R$n$, see text) are shown.  Panel a: R16, initial velocities
differ by $\Delta v=0.06$ km s$^{-1}$ or initial positions 
displaced by 0.015 km from the standard case.
Panel b: R32, initial velocities
differ by $\Delta v=0.06$ km s$^{-1}$ or initial positions 
displaced by 0.0075 km from the standard case.
Panel c: R57, initial positions displaced by 0.0044 km from 
the standard case.  }
\end{figure*}

\begin{figure*}
\centerline{\includegraphics[width=12cm]{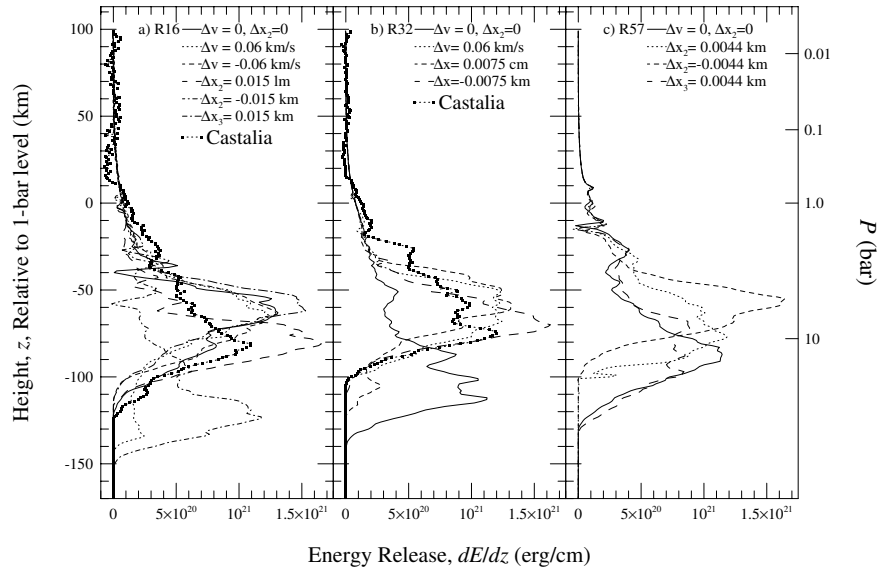}}
\caption{\label{figedeplin}
Same as Fig.\ \ref{figedeplog}, but now with a linear scale
on the horizontal axis
that emphasizes the differences in energy deposition among the runs.  Also
included among the plotted runs are two calculations (at resolutions R16
and R32) of impactors shaped like the asteroid 4769 Castalia.  Both
these runs fall among the main group of runs at each resolution,
indicating that initial macroscopic impactor shape does not strongly
affect energy-deposition behavior. }
\end{figure*}

Our main results are presented in Figs.\ \ref{figedeplog} and
\ref{figedeplin}, which show the profile of
kinetic energy deposition ($dE/dz$) for a number of realizations of the 
standard case described above.  Figs.\ \ref{figedeplog} and \ref{figedeplin}
show much the same
information, plotted in two different ways to emphasize two different points.
Each panel shows profiles generated
by calculations with very slight differences in initial conditions, as
described above.  The panels show calculations done 
at different resolutions: R16, R32, and R57.  
In Fig.\ \ref{figedeplog}, the horizontal scale is logarithmic, to facilitate
a comparison to the results of \citet{MacLowZahnle1994apj} and
\citet{Crawford1996}.  

For the most part previous calculations have been 2D axisymmetric calculations 
at moderately high resolutions (R25) with finite-difference (grid-based) 
hydrocodes.  The exception is the calculation by \citet{TakataEtak1994icsl9}, 
a 3D calculation employing the smooth-particle hydrodynamics (SPH) method.  
Several groups found that a 1-km object penetrated more deeply than we
found, to depths well below $-100$ km, $P>15$ bar 
\citep{BosloughEtal1994grl,CrawfordEtal1994swsl9,TakataEtak1994icsl9,
CrawfordEtal1995ijie}.  \citet{ShuvalovEtal1999ijie} found different results, 
partly due to the different density and structure of the objects they model.  
Some of their calculations were made for an object of $\rho$=0.22 g
cm$^{-3}$.  They also investigated the impacts of objects of
non-uniform density: a dense core ($\rho$ = 1 g cm$^{-3}$) and surrounding 
shell of low density ($\rho$ = 0.1 g cm$^{-3}$), or a low-density object with 
high-density inclusions.  Their objects had the same diameter (1 km) but were 
about 1/3 as massive as our standard object.
\citeauthor{ShuvalovEtal1999ijie} found
a rather broad, double-peaked distribution of energy deposition.  Our
results are most similar to those of \citet{MacLowZahnle1994apj}
and \citet{Crawford1997nyacad}.  The energy-release profile is sharply peaked, 
though not so strongly as found by Mac Low and Zahnle; a small amount of energy
is deposited at levels below 100 km. It is apparent that the sensitivity to 
initial conditions described above also obtains for these simulations.  This is
brought out more strongly in Fig.\ \ref{figedeplin}, in which the
horizontal scale is linear.

An important question about simulations of this type is the degree of 
convergence that they exhibit as a function of numerical parameters 
such as resolution.  In this case convergence means that extrapolation to 
infinite resolution of a series of models would yield a series of results that
converged to the correct answer.  Ideally, the limit of resolution-independent 
results is reached while we are still in the realizable limit of finite 
resolution.  The question in this case is complicated by the sensitivity to 
initial conditions discussed above, since a degree of scatter in the
results is  
introduced, as can be seen in Fig.\ \ref{figedeplin}.  The scatter tends
to  obscure trends in 
the results as a function of resolution.  We must thus compare the
results as a
function of resolution on a statistical basis.  

Also included among the runs plotted in Fig.\ \ref{figedeplin} are two
calculations (at
resolutions R16 and R32) of bolides shaped like the asteroid 4769 Castalia but 
with the same volume and density as our standard case.  These runs are
a test of the influence of the bolide's shape on 
the outcome, the object being in these cases an oblong object with axis ratios 
$2:1:0.8$ and perhaps a representative shape for non-spherical impactors. The 
object in these cases was oriented end-on to the direction of motion. 
The particular model shape was already available for use, having been 
extensively used in the calculations performed by 
\citet{KorycanskyZahnle2003icar}.  The results of the runs of
Castalia-shaped impactors do not appear to be radically different from 
spherical ones.  Large changes in impactor shape do not seem to affect the 
outcome more than tiny changes in the initial position or velocity.  
Presumably, very extreme shapes (e.g., needle-like or plate-shaped objects) 
would have an effect, but moderately oblong shapes are not a strong influence.
For a relatively fragile impactor, it is probably true that the initial shape 
is quickly ``forgotten'' as the incoming object is rapidly deformed by 
aerodynamic forces, and that the same effects that apply to our spherical 
impactors also apply here.  We expect that a sequence of calculations for 
irregular objects, with small changes in initial conditions, would show a 
similar degree of scatter in the results.  \citet{CrawfordEtal1995ijie} also 
simulated the impact of 2D, nonspherical objects (cylinders with
length/diameter
ratios of 1:3 and 3:1) and found a significant but not overwhelming dependence 
on body shape; paradoxically, the 3:1, rod-shaped impactor penetrated
the least
deeply among the three cases tested.

\begin{figure*}
\centerline{\includegraphics[width=14cm]{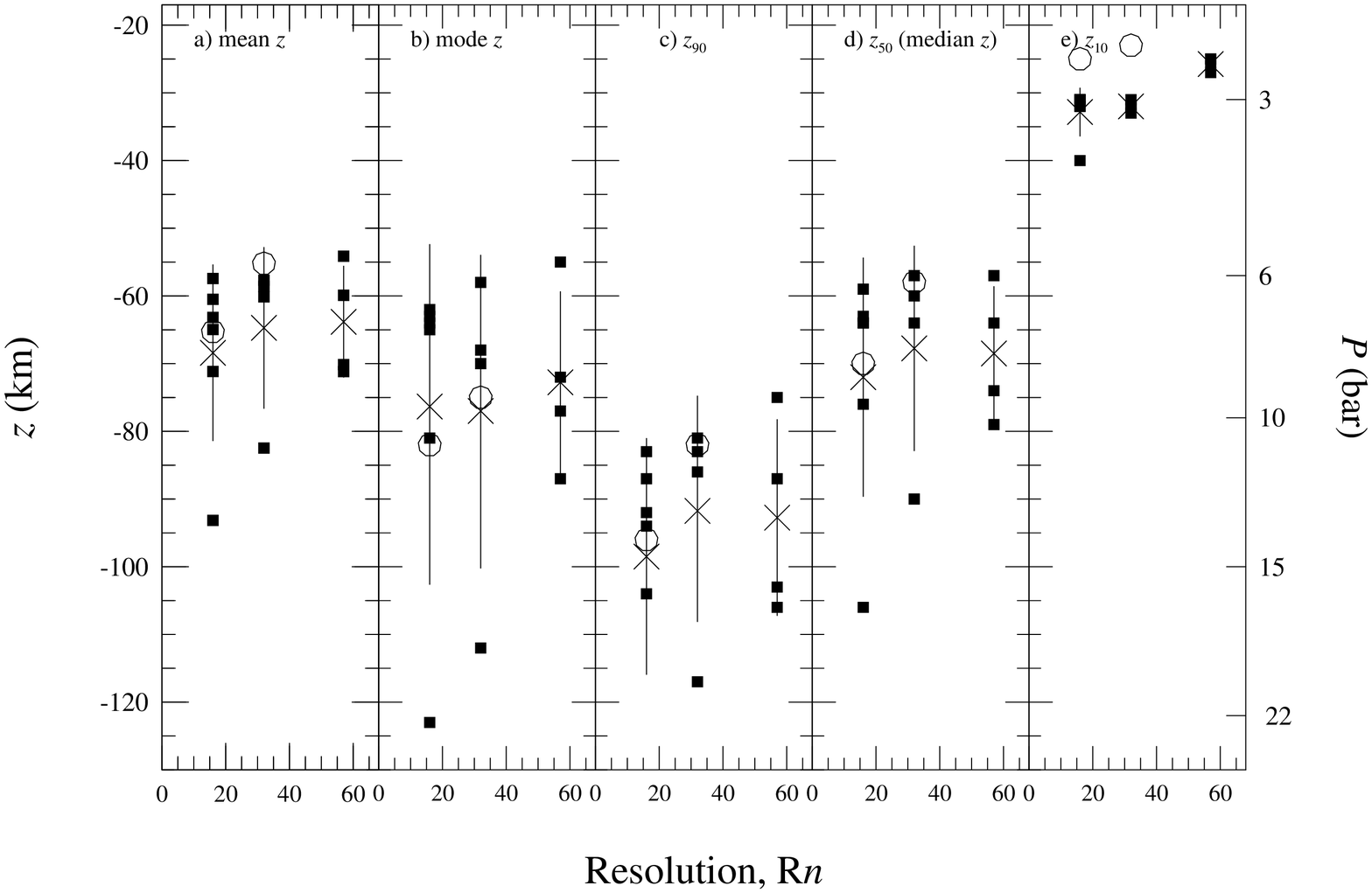}}
\caption{\label{figedepstat}
Measures of energy deposition for the runs shown in Fig.\
\ref{figedeplog}.  Filled 
squares refer to individual runs whose curves are plotted in Fig.\
\ref{figedeplog} at 
resolutions R16, R32, and R57.  The horizontal coordinate in each panel is
the resolution. Crosses are the means of the runs at a given resolution, and
the vertical line extends $\pm$ 1 standard deviation  of the distribution.
The open circles refer to the results of the Castalia-shaped bolides.
Panel a: Mean $z$-value of energy deposition, calculated from the first
moments of the energy-deposition profile.  Panel b: Mode $z$, or the depth
of maximum energy deposition.  Panels c through e, respectively, depth at
which 90\%, 50\%, and 10\% the impact kinetic energy has been deposited.
}
\end{figure*}

Figure \ref{figedepstat} shows various statistical measures
of the energy deposition, in particular several different depths
that characterize the results.  Included in the plot are the mean depth 
(${\bar z}=\int z (dE/dz) dz/E$) or the first moment of the energy deposition 
curve, the mode depth $z$, (the depth of peak energy deposition), and the 
depths, $z_n$, at which $n$\% of the bolide's energy has been deposited for 
$n=$ 90, 50, and 10 ($z_{50}$ is the median depth).  The mean and median depths 
are quite similar despite the skewness of the deposition profiles, while the 
peak energy-deposition depths $z$ are somewhat deeper.  The trend (if any) of 
these measures as a function of resolution is weaker than the amount of scatter,
as seen by the standard deviation.  In particular the R57 runs give 
approximately the same results as the ones at lower resolution.  
Statistical tests ($t$- and $F$-tests) applied to
the various measures of depth for R16 and R57 find that the differences 
between them are not statistically significant.  However, the variance of the
R16 results is $3-4\times$ larger than those of R57, and if the same results 
persisted after $\sim 10$ more cases were run, a significant result might 
emerge.  That is, it is possible that the amount of scatter in the results is 
smaller for high resolution runs.
We also note that the results for the Castalia runs do appear to result in
systematically slightly shallower depths than the means at R16 and R32.
However, only one Castalia run was done for R16 and R32, so the apparent
results may not be significant.

We also examined the spatially-resolved, integrated fluxes $\mu(z;x,y)$ 
described in the previous section.  These provide clues as to exactly how an 
impact proceeds, in terms of understanding the processes of fragmentation,
ablation, and impactor spreading due to aerodynamic forces (pressure 
gradients).  The last process has been denoted ``pancaking'' 
\citep{Zahnle1992jgr, ChybaThomasZahnle1993natu} and has been modeled
semi-analytically and applied to SL9 impacts by several groups
\citep{ChevalierSarazin1994apj, ZahnleMacLow1994icar, MacLowZahnle1994apj,
FieldFerrara1995apj,Crawford1997nyacad}.   We will not make such a model in 
this paper, but simply discuss the hydrodynamical results.  

\begin{figure*}
\centerline{\includegraphics[width=14cm]{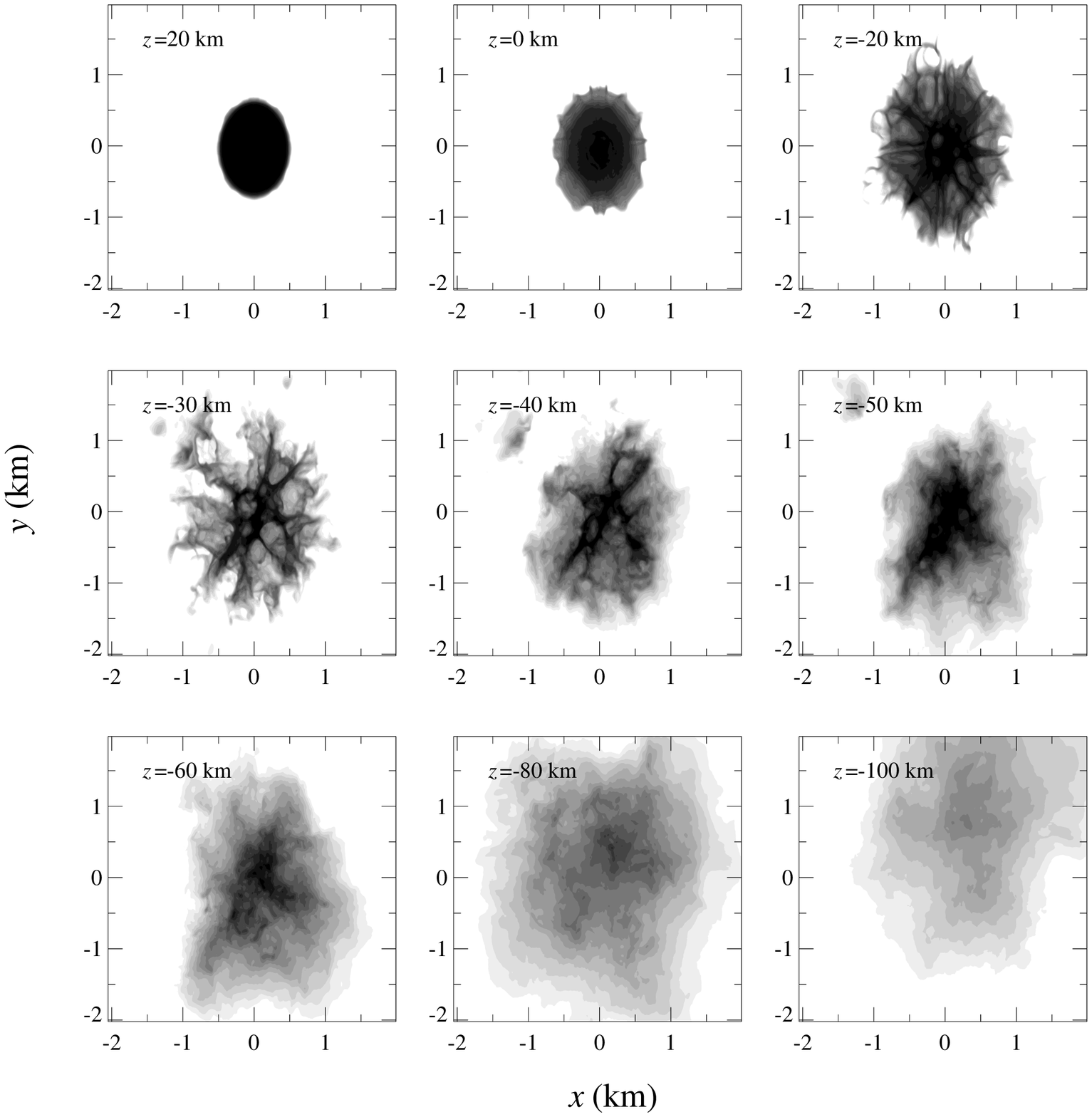}}
\caption{\label{fighorslice}
Time-integrated mass fluxes $\mu(z;x,y)$ as defined in Eq.\
\ref{mudef} for the R57 run at various heights $z$ relative to the 1-bar
level. The fluxes are plotted on a logarithmic gray scale for values of
$\mu > 5\times 10^2$ g cm$^{-2}$ and the darkest shade (black) corresponds
to $\mu > 5\times 10^4$ g cm$^{-2}$.  Only a grid subsection is shown.
}
\end{figure*}

Figure \ref{fighorslice} shows the time-integrated flux $\mu(z;x,y)$
calculated at various
heights $z$ relative to the 1-bar level for the R57 run.  The flux is plotted 
on a logarithmic gray scale  
for $\mu > 5\times 10^2$ g cm$^{-2}$, which emphasizes relatively
small values of $\mu$.  Values of  $5\times 10^4$ g cm$^{-2}$ 
and above are shown in the deepest shade (black).  (For comparison,
note that a 
column 1 km high of $\rho=0.6$ g cm$^{-3}$ has
a surface density of $6\times 10^4$ g cm$^{-2}$.)  Only the inner 
$4\times 4$ km region of the projected grid is shown.  
Due to the zenith angle
of the impact ($\approx 44^{\circ}$) the ``footprint'' of the impact is
elongated in the $+y$ direction in the plots; the effect is most noticeable  
in the $z=20$ km plot, in which the bolide is not yet strongly affected by
the atmosphere. 

The object is torn apart quite high in the atmosphere 
(approximately between $z=-20$ and $-50$ km) compared to the region
below -50 km, where most of the energy is deposited.  Despite the
explosion-like character of the impact, the crushed bolide has enough inertia 
to carry it down another scale height before it stops and deposits its kinetic 
energy.  The same behavior was noted by \citet{ShuvalovEtal1999ijie} in
their calculations.

\begin{figure*}
\centerline{\includegraphics[width=14cm]{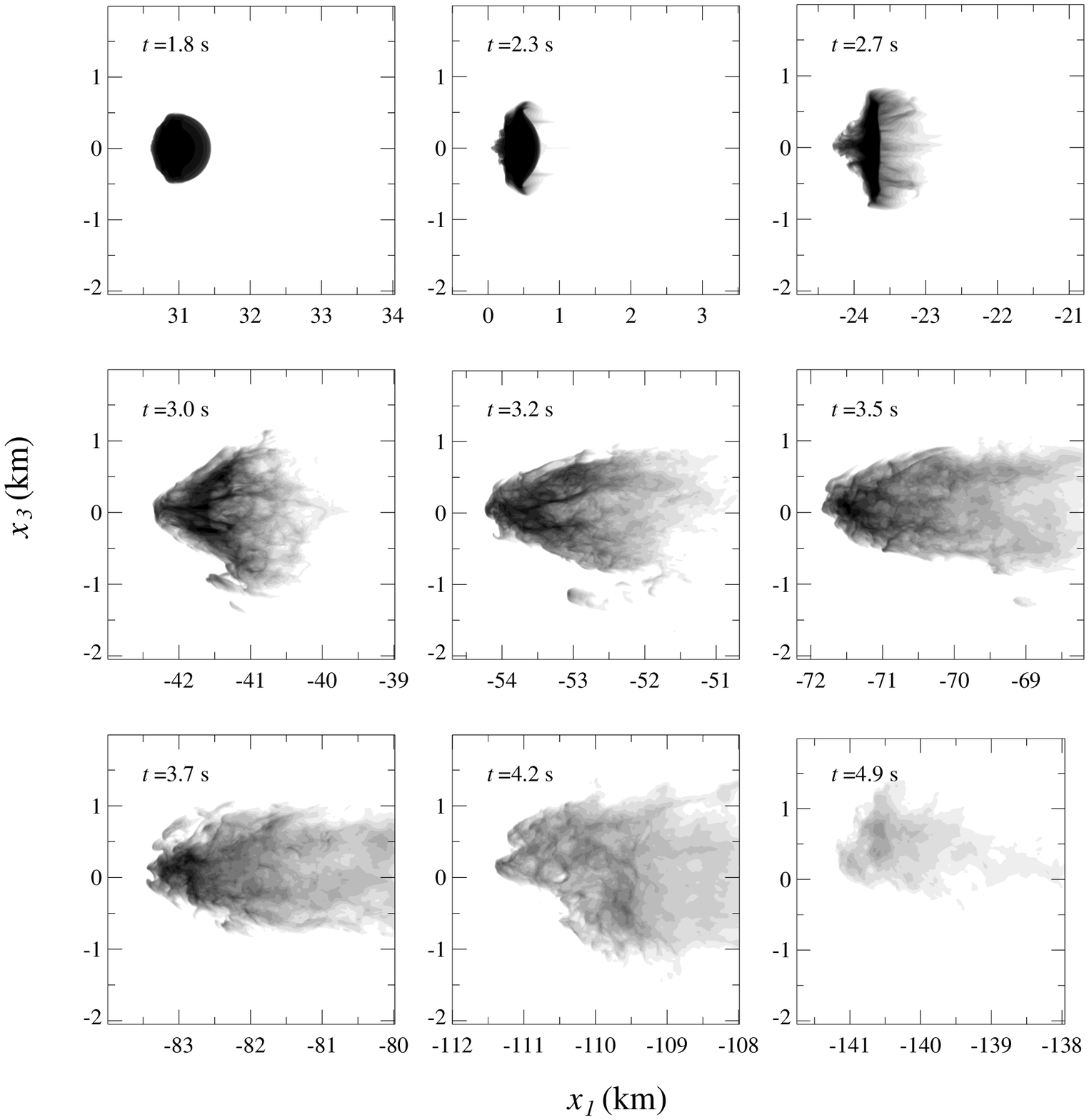}}
\caption{\label{figvertslice}
Column densities $\sigma$ in the $x_1-x_3$ plane (horizontal and
vertical, respectively, for one R57 run.
The times and heights roughly correspond  to the passage of the bolide
through the $z$ planes given in the same panels in Fig.\ \ref{fighorslice}.
Shading is the same as for Fig.\ \ref{fighorslice}, substituting
$\sigma$ for $\mu$.  Only a grid subsection is shown.
}
\end{figure*}

In Fig.\ \ref{figvertslice} we show a similar quantity to $\mu$ for
the R57 run, namely the column density 
$\sigma$ in the ($x_1,x_3$) plane, or a ``side view'' of the impactor generated
by integrating the density of impactor material in the $x_2$ direction:
\begin{equation}
	\sigma(x_1,x_3) = \int \rho C dx_2.
\end{equation}
Note that $\sigma$ is not time-integrated; we show particular instants in the
calculation when the bolide is passing through $z$ levels that are
approximately
the same as those shown in Fig.\ \ref{fighorslice}.  Note also that
Fig.\ \ref{figvertslice} is plotted 
in grid coordinates ($x_1$, $x_3$), which are related to $z$ by 
Eq.\ \ref{coord}.  Figure \ref{figvertslice} shows the compression and 
disruption of the bolide from a different point of view.  The most interesting
part of Fig.\ \ref{figvertslice} is the initial compression of the
bolide seen at 
$t=2.7$ s, followed by the shredding and sweeping back of  
material for $t > 3$ s.  As noted above for Fig.\ \ref{fighorslice},
the lateral spreading 
of the impactor during its initial compression is not larger than a factor
of 2 or so.   

A notable feature is the character of
the impactor disruption.  The impactor is apparently shredded and crushed,
but does not seem to fragment into large pieces that separate at
significant velocity.  Non-axisymmetric filamentary structures develop and
then expand into a cloud of material.
This behavior is different from what has been seen
in calculations of asteroid impacts into the atmosphere of Venus
\citep{KorycanskyEtal2002icar,KorycanskyZahnle2003icar}
and inferred from craters on Venus \citep{KorycanskyZahnle2004icar}
and the Earth \citep{PasseyMelosh1980icar}.  Other calculations at lower
resolution (R32 and R16) show similar behavior.  Given that the same code
was used for Venus calculations \citep{KorycanskyZahnle2003icar} and the
calculations are in many ways very similar, we conclude
that the material of the bolide (porous ice vs. rock) and its
compressibility must control the
character of impactor break-up in these physical situations.

\begin{figure*}
\centerline{\includegraphics[width=14cm]{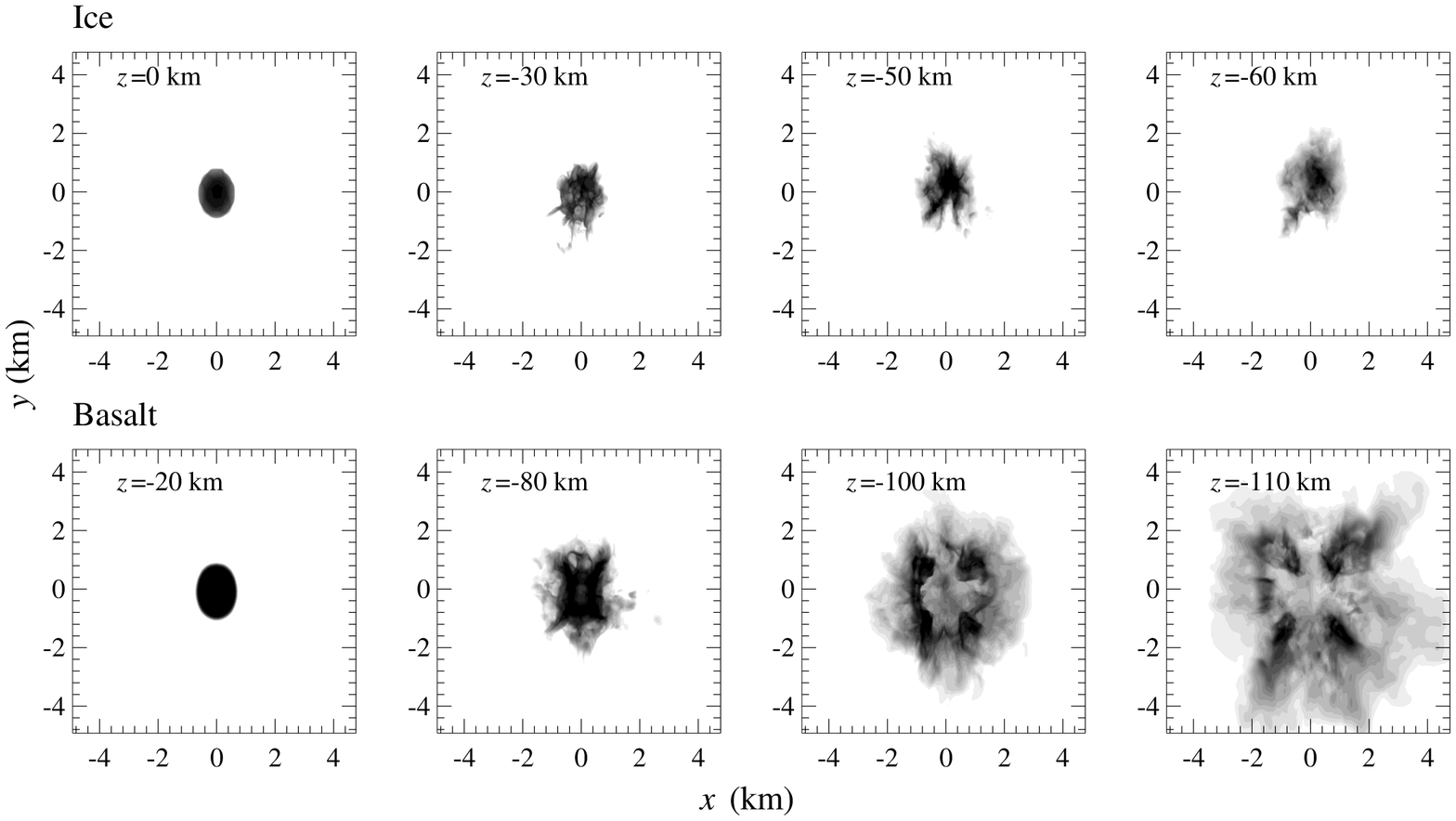}}
\caption{\label{fighorcomp}
Time-integrated mass fluxes $\mu(z;x,y)$ as defined in Eq.\
\ref{mudef} for two R32 runs at various heights, $z$, relative to the 1-bar
level.  The top row is an ice impactor and the bottom row is an impactor of
non-porous basalt ($\rho=2.7$ g cm$^{-3}$).  
Shading is the same as for Fig.\ \ref{fighorslice}.  Only a grid
subsection is shown, but it is larger than that in Fig.\
\ref{fighorslice}.
}
\end{figure*}

\begin{figure*}
\centerline{\includegraphics[width=14cm]{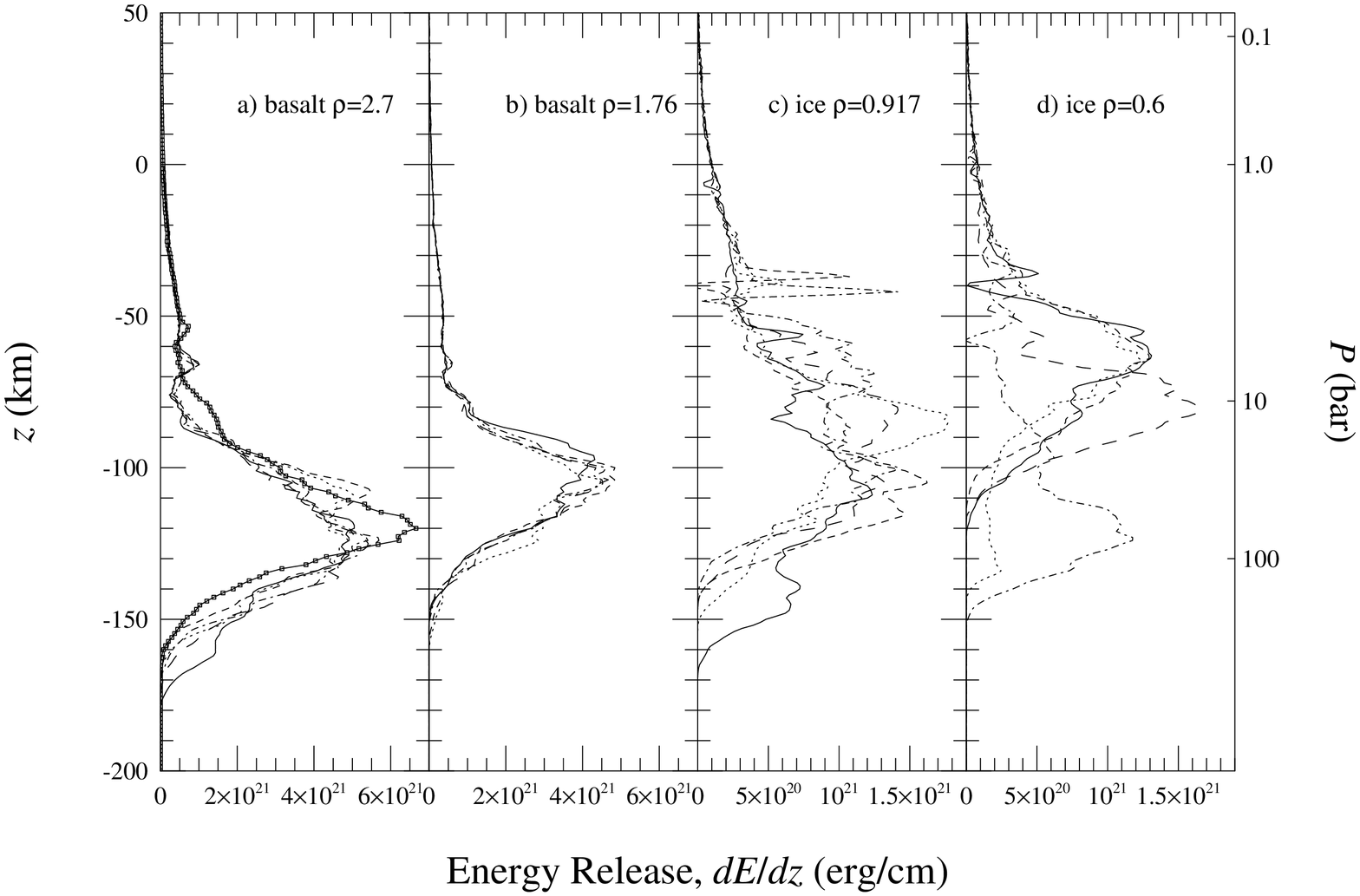}}
\caption{\label{figedep}
Energy-deposition curves for basalt {\vs} ice and porous {\vs}
non-porous 1-km diameter objects.  Five R16 calculations with
differences of 0.015 km in initial  
$x_2$ and $x_3$ positions are shown in each panel.  
Panels a and b are basalt and panels c and d
are ice; panels b and d are for impactors of 35\% initial porosity.
Note the difference in horizontal scale for $dE/dz$ between panels a
and b
and panels c and d.  Panel d shows the same runs those in Figs.\
\ref{figedeplog} and \ref{figedeplin}.
}
\end{figure*}

To test this
idea, we ran an R32 simulation of a 1-km spherical impactor
of non-porous basalt ($\rho=2.7$ g cm$^{-3}$) with
otherwise identical conditions to our standard case.  The results
are shown in Fig.\ \ref{fighorcomp}, where we compare an R32
porous-ice calculation 
(top row) with the basalt impactor (bottom row) at selected heights in the
atmosphere.  Because the basalt bolide is $\sim 5$ times as massive as the ice 
one, it penetrates more deeply, as reflected in the choice  of heights in 
Fig.\ \ref{fighorcomp}.  Of more interest  is the impactor breakup:
the basalt object appears  
to break into several fragments and spread out (pancake) considerably more 
than does the ice impactor, whose degree of pancaking is rather modest, less
than a factor of two in radius.  However, to assess the degree to which the
pancake model does or does not match the behavior seen in Figs.\
\ref{fighorslice} -- \ref{fighorcomp} requires
quantitative modeling that we postpone to future work.

Fig.\ \ref{figedep} shows 
an additional exploration of the possible outcomes due to differing
substances.  We ran five R16 impact calculations for 1-km diameter impactors
of non-porous ice and porous and non-porous basalt.  
The differing impactor masses largely account for 
the variation of average penetration depth per impactor type.
However, the differences in the shapes and the level of variation
among different trials is unexpected. 
The difference in outcomes between ice and basalt is marked; the icy impactors 
show much greater variation in energy deposition (and a much greater overall
spread in height) than do the basalt impactors. Differences in initial porosity
seem to have little effect compared with the difference in material.
Presumably this is due to differences in coefficients in the EOS.  We speculate
that the greater stiffness of the basalt EOS results in greater pancaking
(as in Fig.\ \ref{fighorcomp}) and (somewhat paradoxically) less
variation in the disruption 
and energy deposition as a result.  One way to examine the question is to run
models with simplified and artificially stiffened or softened equations 
of state.  Understanding this result would reveal fundamental impact
physics and may enable simplified models
of atmospheric impacts that do not require extensive hydrodynamic simulation.
We hope to address this question in more detail in the future.  

\begin{figure*}
\centerline{\includegraphics[width=14cm]{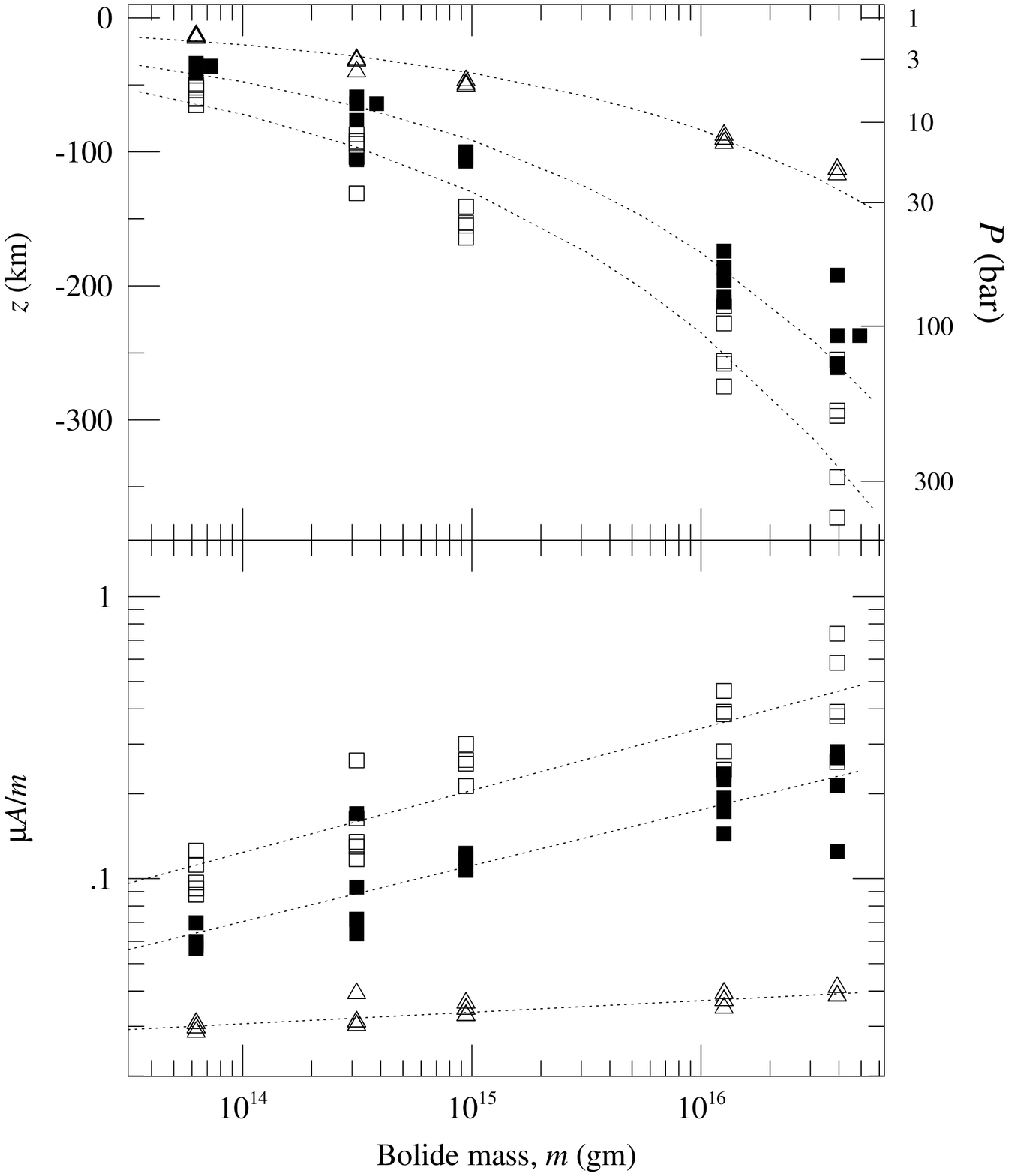}}
\caption{\label{figpenmass}
Depth of bolide penetration into the Jovian atmosphere as a function 
of bolide mass.  Top panel: Depths of energy deposition $z_{10}$, $z_{50}$
and $z_{90}$ for five R16 calculations with tiny perturbations to
their initial positions.  
The dashed lines are fits to the results from Eq.\ \ref{eq:mintp}.
Open triangles are $z_{10}$, filled squares are $z_{50}$, and
open squares are $z_{90}$.
Impactor diameters are 0.584, 1.0, 1.44, 3.42, and 5 km.  The dashed line
is a fit to the results. Bottom panel: Atmosphere column times initial
bolide cross-section, $\mu A$, normalized to bolide mass, as a function
of bolide mass, for 10\%, 50\%, and 90\% levels of energy deposition.  
The dashed lines are fits to the results from Eq.\ \ref{eq:mintp}.
Open triangles are $\mu_{10}A/m$, filled squares are $\mu_{50}A/m$, and
open squares are $\mu_{90}A/m$. }
\end{figure*}

As noted above, we have also run simulations of the impact of bolides of
differing masses corresponding to masses $0.2\times$, $3\times$, $40\times$,
and $125\times$ that of the standard case.  
The corresponding diameters are 0.584, 1.44, 3.42, and 5 km.  We ran 5 cases 
of each diameter at R16 resolution, perturbing the initial positions
in $x_2$ and $x_3$ by one-half grid cell.  The results are shown in
Fig.\ \ref{figpenmass},
where we plot the median depths of energy deposition $z_{10}$, $z_{50}$,
and $z_{90}$ in the 
top panel as a function of bolide mass. The bottom panel shows the same
result, but now we plot the corresponding atmosphere columns 
$\mu_{10,50,90}=\int_{z_{10,50,90}}^{\infty} \rho~dz$ 
times the initial bolide cross-section $A=\pi r^2$, normalized by the 
bolide mass $m$.  Least-squares fits for $z$ and $\mu$ are:
\begin{eqnarray}
z_{10} &=& -9.50\times 10^{-4}(m/{\rm g} )^{0.309} ~{\rm km},\cr
z_{50} &=& -5.19\times 10^{-3}(m/{\rm g} )^{0.283} ~{\rm km},\cr
z_{90} &=& -1.82\times 10^{-2}(m/{\rm g} )^{0.257} ~{\rm km},\cr
{\mu_{10}A\over m} &=& 8.16\times 10^{-3}(m / {\rm g} )^{0.041}, \cr
{\mu_{50}A\over m} &=& 1.20\times 10^{-4}(m / {\rm g} )^{0.199}, \cr
{\mu_{90}A\over m} &=& 1.03\times 10^{-4}(m / {\rm g} )^{0.220}. 
\label{eq:mintp}
\end{eqnarray}
Note that the same sensitivity to initial conditions appears to obtain across
a 600-fold range in impactor mass. 
We expect that impactors would in general penetrate to column depths 
equivalent to their mass, i.e., $\mu A\sim m$, as was found roughly to be the 
case by \citet{KorycanskyZahnle2003icar} for impacts into the Venusian 
atmosphere.  Thus, we
would expect $\mu A\propto m$.  As seen in Eq.\ \ref{eq:mintp} there
is a weak dependence on impactor mass. More massive impactors penetrate
somewhat more deeply than would be expected from a strictly proportional 
relation.  

\section{CONCLUSIONS}
\label{conclsec}

To gain understanding of the SL9 impacts, we have carried out a number
of 3D, hydrodynamic simulations of the impact of porous ice comets
into the atmosphere of Jupiter.  We employed the
numerical hydrodynamics code ZEUS-MP, with some modifications to track the 
comet material (ice), its equation of state, and the degree of porosity, if 
present.  Calculations were carried out at three different resolutions R$n$, 
where $n$ is the number of resolution elements across the bolide radius (for
spherical impactors): R16, R32, and R57.  We have paid special attention to 
the profile of energy deposition in the atmosphere, as a measure of how 
deeply the bolide penetrated and for comparison with previous (mostly 
2D) simulations.  We carried out several calculations of a 
``standard case'' (a 1-km-diameter, porous, ice comet with 
$\rho=0.6$ g cm$^{-3}$ and initial velocity like that of the SL9
impactors) with tiny variations in initial velocity
or shifts in cross-wise position on the computational grid, in order to 
test for sensitivity of the results to initial conditions, and to sample
the distribution of results if the sensitivity were present.  Such multiple
calculations were carried out at all three of our resolutions to see if there
was a convergence trend in the results.  Two low and medium resolution
calculations were also done of an impactor in the shape of 
the asteroid 4769 Castalia to see if there were noticeable differences for
a non-spherical impactor.

Energy-deposition profiles were fairly similar to those found for 
2D calculations such as those done by 
\citet{MacLowZahnle1994apj}, though they were slightly less sharply
peaked.  The median depth of energy deposition was $\approx 70\pm14$
km
below the 1-bar level, at an atmospheric pressure of $\approx $10 bars. 
The aforementioned sensitivity of the results to small changes in initial 
conditions produced significant variations in the energy deposition
profiles, and the error bar just given refers to standard deviation
of the individual profiles.  Comparing the results of calculations at different
resolutions shows very little trend in the results, compared to their
scatter. This suggests that, for the purpose of determining  the depth of 
penetration of impactors, relatively low resolution (R16) is sufficient. 

We have visualized some of the calculations to learn about the impact
process and how a bolide
responds to aerodynamic forces.  The pictures we see
are consistent with  the ``pancake model'' of  \citet{Zahnle1992jgr} and 
\citet{ChybaThomasZahnle1993natu}.  The impactor is flattened quite
strongly at early times,
but the extent of radial spreading was no more than a factor of two
in radius.  Shortly thereafter, the impactor develops non-axisymmetric
structures and shreds into filamentary structures before coming apart 
completely.  Material is blown back and outward as ablation proceeds until the
impactor material expands into a cloud that slows down and deposits its
kinetic energy into the atmosphere.  The disruption takes place at a 
considerably shallower depth (at $\sim -40$ km) than the peak deposition of 
energy; the broken-up impactor material has sufficient inertia to be carried 
downward a significant distance ($\sim 1$ scale height or more) before being 
stopped.

A set of ``low-resolution'' (R16) runs of impactors over a 600-fold range
in mass (corresponding to diameters $0.584 < d < 5$ km) produced median depths 
of energy deposition ranging from 35 km to $\sim 250$ km below the 1-bar level.
Scaling the results by the amount of atmospheric mass intercepted by
the bolide showed a weak dependence on impactor mass, with more massive
bolides penetrating slightly more deeply than predicted by a linear relation
between intercepted atmospheric column and bolide mass.

Future work will extend these results in a number of directions.  We
will explore the parameter space of impactor mass and composition. 
High-resolution calculations will also serve as the basis for new models of 
the impactor plume and splashback that generated the greater part of the
SL9 phenomena observed from Earth.  The results of these calculations will also
serve as input for simplified, semi-analytic, general models of atmospheric 
impacts for diverse situations such as impacts into the atmospheres of 
Earth, Venus, and Titan.   

\acknowledgments

We thank K.\ Zahnle for useful discussions, and the referee, M.-M.\
Mac Low, for helpful suggestions.  This material is based upon work
supported by the National Science Foundation under Grant No.\ 0307638
and upon work supported by the National Aeronautics and Space
Administration under Grant No.\ NNG04GQ35G issued through the Science
Mission Directorate.

\bibliography{ms}

\begin{thebibliography}{53}
\expandafter\ifx\csname natexlab\endcsname\relax\def\natexlab#1{#1}\fi

\bibitem[{{Ahrens} {et~al.}(1994{\natexlab{a}}){Ahrens}, {Takata}, {O'Keefe},
  \& {Orton}}]{AhrensEtal1994grlsl9imp}
{Ahrens}, T.~J., {Takata}, T., {O'Keefe}, J.~D., \& {Orton}, G.~S.
  1994{\natexlab{a}}, \grl, 21, 1087

\bibitem[{{Ahrens} {et~al.}(1994{\natexlab{b}}){Ahrens}, {Takata}, {O'Keefe},
  \& {Orton}}]{AhrensEtal1994grlsl9rad}
---. 1994{\natexlab{b}}, \grl, 21, 1551

\bibitem[{{Asphaug} \& {Benz}(1994)}]{AsphaugBenz1994natsl9}
{Asphaug}, E. \& {Benz}, W. 1994, \nat, 370, 120

\bibitem[{{Asphaug} \& {Benz}(1996)}]{AsphaugBenz1996icsl9}
---. 1996, Icarus, 121, 225

\bibitem[{{Banfield} {et~al.}(1996){Banfield}, {Gierasch}, {Squyres},
  {Nicholson}, {Conrath}, \& {Matthews}}]{BanfieldEtal1996icsl9}
{Banfield}, D., {Gierasch}, P.~J., {Squyres}, S.~W., {Nicholson}, P.~D.,
  {Conrath}, B.~J., \& {Matthews}, K. 1996, Icarus, 121, 389

\bibitem[{{Benz} \& {Asphaug}(1999)}]{BenzAsphaug1999icar}
{Benz}, W. \& {Asphaug}, E. 1999, Icarus, 142, 5

\bibitem[{{B{\'e}zard} {et~al.}(2002){B{\'e}zard}, {Lellouch}, {Strobel},
  {Maillard}, \& {Drossart}}]{BezardEtal2002icsl9co}
{B{\'e}zard}, B., {Lellouch}, E., {Strobel}, D., {Maillard}, J.-P., \&
  {Drossart}, P. 2002, Icarus, 159, 95

\bibitem[{{Boslough} {et~al.}(1994){Boslough}, {Crawford}, {Robinson}, \&
  {Trucano}}]{BosloughEtal1994grl}
{Boslough}, M.~B., {Crawford}, D.~A., {Robinson}, A.~C., \& {Trucano}, T.~G.
  1994, \grl, 21, 1555

\bibitem[{{Boslough} {et~al.}(1995){Boslough}, {Crawford}, {Trucano}, \&
  {Robinson}}]{BosloughEtal1995grlsl9}
{Boslough}, M.~B., {Crawford}, D.~A., {Trucano}, T.~G., \& {Robinson}, A.~C.
  1995, \grl, 22, 1821

\bibitem[{{Boslough} \& {Gladstone}(1997)}]{BosloughEtal1997grldayglow}
{Boslough}, M.~B.~E. \& {Gladstone}, G.~R. 1997, \grl, 24, 3117

\bibitem[{{Brecht} {et~al.}(2001){Brecht}, {de Pater}, {Larson}, \&
  {Pesses}}]{BrechtEtal2001icsl9}
{Brecht}, S.~H., {de Pater}, I., {Larson}, D.~J., \& {Pesses}, M.~E. 2001,
  Icarus, 151, 25

\bibitem[{{Brecht} {et~al.}(1995){Brecht}, {Pesses}, {Lyon}, {Gladd}, \&
  {McDonald}}]{BrechtEtal1995grlsl9}
{Brecht}, S.~H., {Pesses}, M., {Lyon}, J.~G., {Gladd}, N.~T., \& {McDonald},
  S.~W. 1995, \grl, 22, 1805

\bibitem[{{Chevalier} \& {Sarazin}(1994)}]{ChevalierSarazin1994apj}
{Chevalier}, R.~A. \& {Sarazin}, C.~L. 1994, \apj, 429, 863

\bibitem[{{Chodas} \& {Yeomans}(1996)}]{ChodasYeomans1996}
{Chodas}, P.~W. \& {Yeomans}, D.~K. 1996, in IAU Colloq. 156: The Collision of
  Comet Shoemaker-Levy 9 and Jupiter, 1--30

\bibitem[{{Chyba} {et~al.}(1993){Chyba}, {Thomas}, \&
  {Zahnle}}]{ChybaThomasZahnle1993natu}
{Chyba}, C.~F., {Thomas}, P.~J., \& {Zahnle}, K.~J. 1993, \nat, 361, 40

\bibitem[{{Crawford}(1996)}]{Crawford1996}
{Crawford}, D.~A. 1996, in IAU Colloq. 156: The Collision of Comet
  Shoemaker-Levy 9 and Jupiter, 133--156

\bibitem[{{Crawford}(1997)}]{Crawford1997nyacad}
{Crawford}, D.~A. 1997, in Near-Earth Objects, 155--+

\bibitem[{{Crawford} {et~al.}(1994){Crawford}, {Boslough}, {Trucano}, \&
  {Robinson}}]{CrawfordEtal1994swsl9}
{Crawford}, D.~A., {Boslough}, M.~B., {Trucano}, T.~G., \& {Robinson}, A.~C.
  1994, Shock Waves, 4, 47

\bibitem[{{Crawford} {et~al.}(1995){Crawford}, {Boslough}, {Trucano}, \&
  {Robinson}}]{CrawfordEtal1995ijie}
---. 1995, International Journal of Impact Engineering, 17, 253

\bibitem[{{Deming} \& {Harrington}(2001)}]{DemingHarrington2001apjsl9ii}
{Deming}, D. \& {Harrington}, J. 2001, \apj, 561, 468

\bibitem[{{Field} \& {Ferrara}(1995)}]{FieldFerrara1995apj}
{Field}, G.~B. \& {Ferrara}, A. 1995, \apj, 438, 957

\bibitem[{{Gryaznov} {et~al.}(1994){Gryaznov}, {Ivanov}, {Ivlev}, {Klumov},
  {Utyuzhnikov}, \& {Fortov}}]{GryaznovEtal1994emp}
{Gryaznov}, V.~K., {Ivanov}, B.~A., {Ivlev}, A.~B., {Klumov}, B.~A.,
  {Utyuzhnikov}, S.~V., \& {Fortov}, V.~E. 1994, Earth Moon and Planets, 66, 99

\bibitem[{{Hammel} {et~al.}(1995){Hammel}, {Beebe}, {Ingersoll}, {Orton},
  {Mills}, {Simon}, {Chodas}, {Clarke}, {de Jong}, {Dowling}, {Harrington},
  {Huber}, {Karkoschka}, {Santori}, {Tiogo}, {Yeomans}, \&
  {West}}]{HammelEtal1995scisl9}
{Hammel}, H.~B., {Beebe}, R.~F., {Ingersoll}, A.~P., {Orton}, G.~S., {Mills},
  J.~R., {Simon}, A.~A., {Chodas}, P., {Clarke}, J.~T., {de Jong}, E.,
  {Dowling}, T.~E., {Harrington}, J., {Huber}, L.~F., {Karkoschka}, E.,
  {Santori}, C.~M., {Tiogo}, A., {Yeomans}, D., \& {West}, R.~A. 1995, Science,
  267, 1288

\bibitem[{{Harrington} {et~al.}(2004){Harrington}, {de Pater}, {Brecht},
  {Deming}, {Meadows}, {Zahnle}, \& {Nicholson}}]{HarringtonEtal2004jupbksl9}
{Harrington}, J., {de Pater}, I., {Brecht}, S.~H., {Deming}, D., {Meadows},
  V.~S., {Zahnle}, K., \& {Nicholson}, P.~D. 2004, in Jupiter: The Planet,
  Satellites \& Magnetosphere, ed. F.~{Bagenal}, T.~E. {Dowling}, \&
  W.~{McKinnon} (Cambridge, UK: Cambridge University Press), 159--184

\bibitem[{{Harrington} \& {Deming}(2001)}]{HarringtonDeming2001apjsl9i}
{Harrington}, J. \& {Deming}, D. 2001, \apj, 561, 455

\bibitem[{{Herrmann}(1969)}]{Herrmann1969}
{Herrmann}, W. 1969, J. Appl. Phys., 40, 2490

\bibitem[{{Korycansky} \& {Zahnle}(2003)}]{KorycanskyZahnle2003icar}
{Korycansky}, D.~G. \& {Zahnle}, K.~J. 2003, Icarus, 161, 244

\bibitem[{{Korycansky} \& {Zahnle}(2004)}]{KorycanskyZahnle2004icar}
---. 2004, Icarus, 169, 287

\bibitem[{{Korycansky} {et~al.}(2000){Korycansky}, {Zahnle}, \& {Mac
  Low}}]{KorycanskyEtal2000icar}
{Korycansky}, D.~G., {Zahnle}, K.~J., \& {Mac Low}, M.-M. 2000, Icarus, 146,
  387

\bibitem[{{Korycansky} {et~al.}(2002){Korycansky}, {Zahnle}, \& {Mac
  Low}}]{KorycanskyEtal2002icar}
---. 2002, Icarus, 157, 1

\bibitem[{{Lellouch} {et~al.}(2002){Lellouch}, {B{\' e}zard}, {Moses}, {Davis},
  {Drossart}, {Feuchtgruber}, {Bergin}, {Moreno}, \&
  {Encrenaz}}]{LellouchEtal2002icsl9co2h20}
{Lellouch}, E., {B{\' e}zard}, B., {Moses}, J., {Davis}, G., {Drossart}, P.,
  {Feuchtgruber}, H., {Bergin}, E., {Moreno}, R., \& {Encrenaz}, T. 2002,
  Icarus, 159, 112

\bibitem[{{Mac Low}(1996)}]{MacLow1996}
{Mac Low}, M.-M. 1996, in IAU Colloq. 156: The Collision of Comet
  Shoemaker-Levy 9 and Jupiter, 157--182

\bibitem[{{Mac Low} \& {Zahnle}(1994)}]{MacLowZahnle1994apj}
{Mac Low}, M.-M. \& {Zahnle}, K. 1994, \apjl, 434, L33

\bibitem[{{McGregor} {et~al.}(1996){McGregor}, {Nicholson}, \&
  {Allen}}]{McgregorEtal1996icsl9ring}
{McGregor}, P.~J., {Nicholson}, P.~D., \& {Allen}, M.~G. 1996, Icarus, 121, 361

\bibitem[{{Melosh}(1989)}]{Melosh1989}
{Melosh}, H.~J. 1989, {Impact cratering: A geologic process} (Research
  supported by NASA.~New York, Oxford University Press (Oxford Monographs on
  Geology and Geophysics, No.~11), 1989, 253 p.), 45--+

\bibitem[{{Menikoff} \& {Kober}(1999)}]{MenikoffKober1999}
{Menikoff}, R. \& {Kober}, E. 1999, in APS Topical Conference on Shock
  Compression in Condensed Matter, Snowbird, Utah, June 27 - July 2, 1999.

\bibitem[{{Noll} {et~al.}(1996){Noll}, {Weaver}, \&
  {Feldman}}]{NollEtal1996bksl9iau156}
{Noll}, K.~S., {Weaver}, H.~A., \& {Feldman}, P.~D., eds. 1996, The Collision
  of Comet {Shoemaker-Levy 9} and {Jupiter}, {IAU} Colloquium 156 (Cambridge,
  UK: Cambridge University Press)

\bibitem[{{Norman}(2000)}]{Norman2000}
{Norman}, M.~L. 2000, in Revista Mexicana de Astronomia y Astrofisica
  Conference Series, 66--71

\bibitem[{{Passey} \& {Melosh}(1980)}]{PasseyMelosh1980icar}
{Passey}, Q.~R. \& {Melosh}, H.~J. 1980, Icarus, 42, 211

\bibitem[{{Shoemaker} {et~al.}(1995){Shoemaker}, {Hassig}, \&
  {Roddy}}]{ShoemakerEtal1995grl}
{Shoemaker}, E.~M., {Hassig}, P.~J., \& {Roddy}, D.~J. 1995, \grl, 22, 1825

\bibitem[{{Shuvalov} {et~al.}(1999){Shuvalov}, {Artem'eva}, \&
  Kosarev}]{ShuvalovEtal1999ijie}
{Shuvalov}, V.~V., {Artem'eva}, N.~A., \& Kosarev, I.~B. 1999, Int. J. Impact.
  Eng., 23, 847

\bibitem[{{Shuvalov} {et~al.}(1997){Shuvalov}, {Artem'eva}, {Kosarev},
  Nemtchinov, \& {Trubetskaya}}]{ShuvalovEtal1997sola}
{Shuvalov}, V.~V., {Artem'eva}, N.~A., {Kosarev}, I.~B., Nemtchinov, I.~V., \&
  {Trubetskaya}, I. 1997, Sol. Syst. Res., 31, 393

\bibitem[{{Solem}(1994)}]{Solem1994natsl9}
{Solem}, J.~C. 1994, \nat, 370, 349

\bibitem[{{Solem}(1995)}]{Solem1995aasl9}
---. 1995, \aap, 302, 596

\bibitem[{{Svetsov}(1995)}]{Svetsov1995sola}
{Svetsov}, V.~V. 1995, Sol. Syst. Res., 29, 331

\bibitem[{{Takata} \& {Ahrens}(1997)}]{TakataAhrens1997icsl9}
{Takata}, T. \& {Ahrens}, T.~J. 1997, Icarus, 125, 317

\bibitem[{{Takata} {et~al.}(1994){Takata}, {O'Keefe}, {Ahrens}, \&
  {Orton}}]{TakataEtak1994icsl9}
{Takata}, T., {O'Keefe}, J.~D., {Ahrens}, T.~J., \& {Orton}, G.~S. 1994,
  Icarus, 109, 3

\bibitem[{{Tillotson}(1962)}]{Tillotson1962}
{Tillotson}, J.~H. 1962, General Atomic Report GA-3216

\bibitem[{{West} \& {B{\" o}hnhardt}(1995)}]{WestBoenhardt1995bkesosl9}
{West}, R.~M. \& {B{\" o}hnhardt}, H., eds. 1995, Proceedings of the {European}
  {Shoemaker-Levy 9} Conference, held 13--15 {February} 1995, {ESO} Conference
  and Workshop Proceedings No.~52, European Southern Observatory, Garching bei
  M{\" u}nchen, Germany

\bibitem[{{Yabe} {et~al.}(1995){Yabe}, {Aoki}, {Tajima}, {Xiao}, {Sasaki},
  {Abe}, \& {Watanabe}}]{YabeEtal1995grl}
{Yabe}, T., {Aoki}, T., {Tajima}, M., {Xiao}, F., {Sasaki}, S., {Abe}, Y., \&
  {Watanabe}, J.-I. 1995, \grl, 22, 2429

\bibitem[{{Yabe} {et~al.}(1994){Yabe}, {Xiao}, {Zhang}, {Sasaki}, {Kobayashi},
  \& {Terasawa}}]{YabeEtal1994}
{Yabe}, T., {Xiao}, F., {Zhang}, D.~L., {Sasaki}, S., {Kobayashi}, N., \&
  {Terasawa}, T. 1994, J. Geomag. Geolelectr., 46, 657

\bibitem[{{Zahnle} \& {Mac Low}(1994)}]{ZahnleMacLow1994icar}
{Zahnle}, K. \& {Mac Low}, M.-M. 1994, Icarus, 108, 1

\bibitem[{{Zahnle}(1992)}]{Zahnle1992jgr}
{Zahnle}, K.~J. 1992, \jgr, 97, 10243

\end{thebibliography}

\end{document}